\documentstyle[12pt,epsf,epsfig,wrapfig]{article}
\begin{document}

\begin{tabbing}
\` DOE/ER/40561-289-INT96-00-147
\end{tabbing}

\begin{center}
{\large \bf
Spin Observables in Consistent Relativistic Models of
$\left(e, e^{\prime}p\right)$ and $\left(\gamma, p\right)$
Reactions$^{\ast}$
\\ }
\vspace{5mm}
\underline{J.I. Johansson}$^{1,2}$ and H.S. Sherif$^3$
\\
\vspace{5mm}
{\small\it
(1) Institute for Nuclear Theory, University of Washington, Box 351550,
    Seattle, WA, 98195-1550
\\
(2) Department of Physics, University of Manitoba, Winnipeg, Manitoba, R3T 2N2
\\
(3) Department of Physics, University of Alberta, Edmonton, Alberta, T6G 2J1
\\ }
\end{center}

\begin{center}
ABSTRACT

\vspace{5mm}
\begin{minipage}{130 mm}
\small
Single nucleon knockout reactions must be described in a consistent
framework in order to extract information about nuclear structure
and reaction mechanisms.
Consistent relativistic models for the direct knockout contribution
to the $\left(e, e^{\prime} p\right)$ and
$\left(\gamma, p\right)$ reactions have been developed and used
previously to examine existing momentum distribution and cross section
data for the two reactions$^{1}$.
We present results of calculations of spin observables obtained from
these models of $\left(e, e^{\prime} p\right)$ and
$\left(\gamma, p\right)$ reactions.
For the $\left(e, e^{\prime} p\right)$ reaction we consider the 
importance of the choice of kinematics and show that perpendicular
kinematics result in distributions with relatively large 
cross sections and comparatively large polarizations.
In the $\left(\gamma,p\right)$ reactions the photon asymmetry and 
proton polarization are not very sensitive to changes in 
model ingredients when the incident photon energy is less than 100 $MeV$.
For higher energies these observables show increasing sensitivity
to modifications of both the bound and continuum wave functions.
\end{minipage}
\end{center}

%\section{Introduction}

Relativistic and non-relativistic descriptions of the removal of
a single nucleon from a nucleus using an electromagnetic probe
show intriguing differences.
Spectroscopic factors obtained through relativistic analyses
of $\left(e, e^{\prime} p\right)$ reactions are consistently larger
by 10\% to 20\% than values obtained via non-relativistic
analyses$^{1-4}$.
In addition, relativistic calculations of the direct knockout
contribution to $\left(\gamma, p\right)$ reactions provide a reasonable
description of experimental data over a wide range of nuclear targets
and photon energies$^{1,5}$ while non-relativistic analyses can fall
below the data by more than a factor of five$^{6-8}$.
The non-relativistic analyses suggest a need for large contributions
from meson exchange currents (MEC), while no large MEC
contributions are required in the relativistic analyses.

Clearly we can not currently claim to fully understand the
reaction mechanisms for these reactions in spite of the 
relative simplicity of the electromagnetic probes compared to
strongly interacting probes.
Further experimental guidance is needed, and can be obtained through
measurements of spin dependent observables
which can provide sensitive tests of the various models for
these reactions.

%\section{The Model}

Our models for these reactions retain the fully relativistic
treatment of the amplitudes,
utilizing solutions of the Dirac equation containing strong vector
and scalar potentials to describe the bound and continuum nucleons.
For the $\left(e, e^{\prime} p\right)$ case the amplitude is
calculated in the one photon exchange approximation, while for
$\left(\gamma, p\right)$ we consider only the direct knockout
of the proton$^{9}$.
The spectroscopic factor is determined by scaling our calculations
of the momentum distributions for $\left(e, e^{\prime} p\right)$
reactions to match experimental data.
The same ingredients are then used in our $\left(\gamma, p\right)$
model to perform calculations with {\em no} adjustable
parameters, thus providing {\em predictions} of calculated
observables.
One detail to note is that we do not include full Coulomb 
distortion of the electron wave functions in our model of the
$\left(e, e^{\prime} p\right)$ reaction, but choose to use
the {\em effective momentum approximation}$^{10}$, in which
the magnitude of the electron momentum is modified by the
value of the Coulomb potential at the origin, while its 
direction is unchanged.
This provides a large part of the effect of the Coulomb
distortion on the electron wave functions for the $^{208}Pb$
target for the kinematics considered.
In our model of the $\left(e, e^{\prime} p\right)$ reaction,
the analyzing power due to initially polarized
electrons, and the polarization of the final state electrons
turn out to be less than 1\%.
This is likely due to the lack of a proper inclusion of Coulomb
distortions.

%\section{Spin Observables in $\left(e, e^{\prime} p\right)$}

%
%
% the following 8 cm wide figure will be placed in 
% a 8 cm wide box on the right side of the page 
%
\begin{wrapfigure}{r}{7cm}
\epsfig{figure=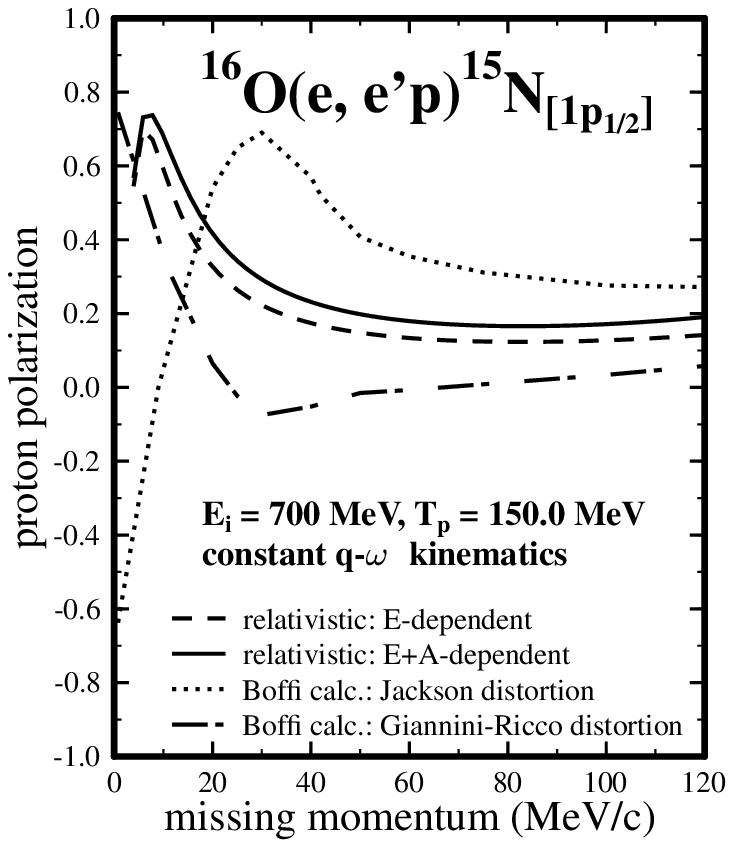,width=7cm}
{\small Figure 1: Proton polarization in relativistic and
non-relativistic models.}
\end{wrapfigure}
The proton polarization obtained in this model for
$\left(e, e^{\prime} p\right)$ should not be sensitive to this 
approximation and we have considered this observable in
three possible kinematic arrangements used to obtain
experimental momentum distributions.
In parallel kinematics the polarization tends to be small, 
showing some structure and an occasional spike but generally
remaining under 20\%.
In both perpendicular and constant $q-\omega$ kinematics, 
the polarization tends to be large and associated
with cross sections large enough to be measurable.
An example is shown in FIG. 1, for the case of an oxygen
target and an incident electron energy of 700 $MeV$.
The proton kinetic energy is close to 150 $MeV$ in constant
$q-\omega$ kinematics.
There are two curves showing relativistic calculations,
(solid and dashed) for two different optical potentials and
a relativistic Dirac-Hartree bound state.
The other two curves (dotted and dash-dotted) are from a
non-relativistic calculation by Boffi {\em et al.}$^{11}$.
Note that at a missing momentum of 30 $MeV/c$ the 
calculated proton polarizations can differ by as much as 0.8.
In this kinematic region the triple differential cross section
is calculated to be about 0.6 $nb / sr sr MeV$ and should be
readily measurable with the facilities now becoming available.

%\section{Spin Observables in $\left(\gamma, p\right)$}

Spin observables for the $\left(\gamma, p\right)$ reaction have
also been studied.
Calculations have been done for both photon asymmetry and proton
polarization and both observables can be sensitive to the
reaction mechanism (i.e. relativistic vs. non-relativistic, or
contributions of MECs, $\Delta$s),
as well as being sensitive to changes in
both bound and continuum wave functions for photon energies
above 100 $MeV$.
A useful approach would be to measure the spin observables
at low photon energies ($E_{\gamma} \approx$ 80-100 $MeV$)
and over a range of target masses, to
try to establish the reaction mechanism in the simplest region.
Then measurements through the $\Delta$-region would be very useful
in understanding the contributions of MECs and isobars to the
reaction mechanism.

%\section{Conclusions}

The relativistic calculations provide a good description of 
$\left(e, e^{\prime} p\right)$ and $\left(\gamma, p\right)$
cross section data even at large missing momenta,
but measurements of spin observables will provide a more
stringent test and would hopefully also clarify the issue of
relativistic versus non-relativistic approaches.
Measurements of proton polarization in the
$\left(e, e^{\prime} p\right)$ reaction will be easiest
in perpendicular or constant $q-\omega$ kinematics, where
large polarizations can be associated with large
cross sections.
Photon asymmetry and proton polarization in the
$\left(\gamma, p\right)$ reaction are both sensitive to the
reaction mechanism as well as the model ingredients.

In short, the reaction mechanism leading to knockout of a
single nucleon from a nucleus by an electromagnetic probe is not
yet clear, and the additional information provided by
measurements of spin dependent observables should strongly
constrain the various models for these reactions.

\vspace{5 mm} 

\noindent $^{\ast}$ Work supported in part by the Natural Sciences and
Engineering Research Council of Canada. One of the authors (H.S.S.)
is grateful to the Theory Group at TRIUMF for their hospitality.

\vspace{0.2cm}
\vfill
{\small\begin{description}

\item{[1]} 
J.I. Johansson, H.S. Sherif and G.M. Lotz, 
Nucl. Phys. {\bf A605} (1996) 517.

\item{[2]} 
M. Hedayati-Poor, J.I. Johansson and H.S. Sherif,
Phys. Rev. C {\bf 51} (1995) 2044.

\item{[3]} 
J.P. McDermott, Phys. Rev. Lett. {\bf 65} (1990) 1991.

\item{[4]} 
J.M. Udias, P. Sarriguren, E. Moya de Guerra,
E. Garrido and J.A. Caballero,
Phys. Rev. C {\bf 48} (1993) 2731.

\item{[5]} 
J.I. Johansson and H.S. Sherif,
to be published.

\item{[6]} 
D.G. Ireland and G. van der Steenhoven,
Phys. Rev. C {\bf  49} (1994) 2182.

\item{[7]}
I. Bobeldijk {\it et al.},
Phys. Lett. B {\bf 356} (1995) 13.

\item{[8]}
K. Mori {\it et al.},
Phys. Rev. C {\bf 51} (1995) 2611.
 
\item{[9]} 
G.M. Lotz and H.S. Sherif, 
Nucl. Phys. {\bf A537} (1992) 285.

\item{[10]}
D.R. Yennie, F.L. Boos, Jr. and D.G. Ravenhall,
Phys. Rev. {\bf  137} (1965) 882.

\item{[11]}
S. Boffi, C. Giusti, F.D. Pacati and M. Radici,
Nucl. Phys. {\bf  A518} (1990) 639.

\end{description}}

\end{document}